\documentclass[aps,prd,twocolumn, groupedaddress]{revtex4-1}
\usepackage[pdftex]{graphicx}
\usepackage{subfigure}
\usepackage{sidecap}

\usepackage{amsmath}
\usepackage{hyperref}
\usepackage{aas_macros}

\graphicspath{{pics/}{plots/}}

\usepackage[pdftex]{color}
\usepackage[pdftex]{graphicx}
\usepackage{amsfonts}
\usepackage{amsmath}
\usepackage{bm}                 % bold math

\newcommand{\sslabel}[1]{\label{sec:#1}}
\newcommand{\elabel}[1]{\label{eqn:#1}}
\newcommand{\flabel}[1]{\label{fig:#1}}

\newcommand{\cref}[1]{\ref{cpt:#1}}

\newcommand{\eref}[1]{\ref{eqn:#1}}
\newcommand{\fref}[1]{\ref{fig:#1}}

\newcommand{\listBegin}{\begin{tabular}{cp{4.5in}}}
\newcommand{\listEnd}{\end{tabular}}

\newcommand{\matrixBegin}[1]{\left[\!\!\left[ \begin{array}{#1}}
\newcommand{\matrixEnd}{\end{array} \right]\!\!\right]}

\newcommand{\beq}[1]{\begin{equation}\elabel{#1}}
\newcommand{\eeq}{\end{equation}}

\newcommand{\beqa}[1]{\begin{eqnarray}\elabel{#1}}
\newcommand{\eeqa}{\end{eqnarray}}

\newcommand{\MEfig}[5]{\begin{figure}[#1] \begin{center}\includegraphics[width=#2\textwidth]{#3}\end{center} \vspace{-10pt}\caption{\flabel{#4}#5}\end{figure}}

\def\({\left(}
\def\){\right)}
\def\[{\left[}
\def\]{\right]}

\def\gw{gravitational wave}
\def\gws{gravitational waves}

\newcommand{\SI}[2]{\ensuremath{#1\,\rm #2}}

\definecolor{spring}{rgb}{0.7,0.9,0.7}
\definecolor{brick}{rgb}{0.7,0.2,0.1}
\definecolor{redHL}{rgb}{1.0,0.5,0.5}

\begin{document}
\def\vup{\vspace{-4pt}}

%\preprint{}

\title{Exploring the Sensitivity of Next Generation Gravitational Wave Detectors}



%% LSC authorlist in CQG format
\author{%
B~P~Abbott$^{1}$,  %benjamin.abbott
R~Abbott$^{1}$,  %rich.abbott
T~D~Abbott$^{2}$,  %thomas.abbott
M~R~Abernathy$^{3}$,  %matthew.abernathy
K~Ackley$^{4}$,  %kendall.ackley
C~Adams$^{5}$,  %carl.adams
P~Addesso$^{6}$,  %paolo.addesso
R~X~Adhikari$^{1}$,  %rana.adhikari
V~B~Adya$^{7}$,  %vaishali.adya
C~Affeldt$^{7}$
% ...
}
\medskip
\address {${}^{**}$Deceased, March 2016. ${}^{\dag}$Deceased, May 2015. ${}^{\ddag}$Deceased, March 2015. }% 
\medskip
\address {$^{1}$LIGO, California Institute of Technology, Pasadena, CA 91125, USA }
\address {$^{2}$Louisiana State University, Baton Rouge, LA 70803, USA }
\address {$^{3}$American University, Washington, D.C. 20016, USA }
\address {$^{4}$University of Florida, Gainesville, FL 32611, USA }
\address {$^{5}$LIGO Livingston Observatory, Livingston, LA 70754, USA }
\address {$^{6}$University of Sannio at Benevento, I-82100 Benevento, Italy and INFN, Sezione di Napoli, I-80100 Napoli, Italy }
\address {$^{7}$Albert-Einstein-Institut, Max-Planck-Institut f\"ur Gravi\-ta\-tions\-physik, D-30167 Hannover, Germany }
\address {$^{8}$LIGO, Massachusetts Institute of Technology, Cambridge, MA 02139, USA }
\address {$^{9}$Instituto Nacional de Pesquisas Espaciais, 12227-010 S\~{a}o Jos\'{e} dos Campos, S\~{a}o Paulo, Brazil }
\address {$^{10}$Inter-University Centre for Astronomy and Astrophysics, Pune 411007, India }
\address {$^{11}$International Centre for Theoretical Sciences, Tata Institute of Fundamental Research, Bangalore 560012, India }
\address {$^{12}$University of Wisconsin-Milwaukee, Milwaukee, WI 53201, USA }
\address {$^{13}$Leibniz Universit\"at Hannover, D-30167 Hannover, Germany }
\address {$^{14}$Australian National University, Canberra, Australian Capital Territory 0200, Australia }
\address {$^{15}$The University of Mississippi, University, MS 38677, USA }
\address {$^{16}$California State University Fullerton, Fullerton, CA 92831, USA }
\address {$^{17}$Chennai Mathematical Institute, Chennai 603103, India }
\address {$^{18}$University of Southampton, Southampton SO17 1BJ, United Kingdom }
\address {$^{19}$Universit\"at Hamburg, D-22761 Hamburg, Germany }
\address {$^{20}$Albert-Einstein-Institut, Max-Planck-Institut f\"ur Gravitations\-physik, D-14476 Potsdam-Golm, Germany }
\address {$^{21}$Montana State University, Bozeman, MT 59717, USA }
\address {$^{22}$Syracuse University, Syracuse, NY 13244, USA }
\address {$^{23}$SUPA, University of Glasgow, Glasgow G12 8QQ, United Kingdom }
\address {$^{24}$LIGO Hanford Observatory, Richland, WA 99352, USA }
\address {$^{25}$Columbia University, New York, NY 10027, USA }
\address {$^{26}$Stanford University, Stanford, CA 94305, USA }
\address {$^{27}$University of Birmingham, Birmingham B15 2TT, United Kingdom }
\address {$^{28}$RRCAT, Indore MP 452013, India }
\address {$^{29}$Faculty of Physics, Lomonosov Moscow State University, Moscow 119991, Russia }
\address {$^{30}$SUPA, University of the West of Scotland, Paisley PA1 2BE, United Kingdom }
\address {$^{31}$University of Western Australia, Crawley, Western Australia 6009, Australia }
\address {$^{32}$Washington State University, Pullman, WA 99164, USA }
\address {$^{33}$University of Oregon, Eugene, OR 97403, USA }
\address {$^{34}$Carleton College, Northfield, MN 55057, USA }
\address {$^{35}$University of Maryland, College Park, MD 20742, USA }
\address {$^{36}$Center for Relativistic Astrophysics and School of Physics, Georgia Institute of Technology, Atlanta, GA 30332, USA }
\address {$^{37}$NASA/Goddard Space Flight Center, Greenbelt, MD 20771, USA }
\address {$^{38}$RESCEU, University of Tokyo, Tokyo, 113-0033, Japan. }
\address {$^{39}$Tsinghua University, Beijing 100084, China }
\address {$^{40}$Texas Tech University, Lubbock, TX 79409, USA }
\address {$^{41}$The Pennsylvania State University, University Park, PA 16802, USA }
\address {$^{42}$National Tsing Hua University, Hsinchu City, 30013 Taiwan, Republic of China }
\address {$^{43}$Charles Sturt University, Wagga Wagga, New South Wales 2678, Australia }
\address {$^{44}$West Virginia University, Morgantown, WV 26506, USA }
\address {$^{45}$University of Chicago, Chicago, IL 60637, USA }
\address {$^{46}$Caltech CaRT, Pasadena, CA 91125, USA }
\address {$^{47}$Korea Institute of Science and Technology Information, Daejeon 305-806, Korea }
\address {$^{48}$University of Brussels, Brussels 1050, Belgium }
\address {$^{49}$Sonoma State University, Rohnert Park, CA 94928, USA }
\address {$^{50}$Center for Interdisciplinary Exploration \& Research in Astrophysics (CIERA), Northwestern University, Evanston, IL 60208, USA }
\address {$^{51}$University of Minnesota, Minneapolis, MN 55455, USA }
\address {$^{52}$The University of Melbourne, Parkville, Victoria 3010, Australia }
\address {$^{53}$Institute for Plasma Research, Bhat, Gandhinagar 382428, India }
\address {$^{54}$The University of Sheffield, Sheffield S10 2TN, United Kingdom }
\address {$^{55}$The University of Texas Rio Grande Valley, Brownsville, TX 78520, USA }
\address {$^{56}$Cardiff University, Cardiff CF24 3AA, United Kingdom }
\address {$^{57}$Montclair State University, Montclair, NJ 07043, USA }
\address {$^{58}$MTA E\"otv\"os University, ``Lendulet'' Astrophysics Research Group, Budapest 1117, Hungary }
\address {$^{59}$School of Mathematics, University of Edinburgh, Edinburgh EH9 3FD, United Kingdom }
\address {$^{60}$Indian Institute of Technology, Gandhinagar Ahmedabad Gujarat 382424, India }
\address {$^{61}$University of Szeged, D\'om t\'er 9, Szeged 6720, Hungary }
\address {$^{62}$Embry-Riddle Aeronautical University, Prescott, AZ 86301, USA }
\address {$^{63}$Tata Institute of Fundamental Research, Mumbai 400005, India }
\address {$^{64}$University of Michigan, Ann Arbor, MI 48109, USA }
\address {$^{65}$Rochester Institute of Technology, Rochester, NY 14623, USA }
\address {$^{66}$University of Massachusetts-Amherst, Amherst, MA 01003, USA }
\address {$^{67}$NCSA, University of Illinois at Urbana-Champaign, Urbana, Illinois 61801, USA }
\address {$^{68}$Universitat de les Illes Balears, IAC3---IEEC, E-07122 Palma de Mallorca, Spain }
\address {$^{69}$SUPA, University of Strathclyde, Glasgow G1 1XQ, United Kingdom }
\address {$^{70}$IISER-TVM, CET Campus, Trivandrum Kerala 695016, India }
\address {$^{71}$Canadian Institute for Theoretical Astrophysics, University of Toronto, Toronto, Ontario M5S 3H8, Canada }
\address {$^{72}$Institute of Applied Physics, Nizhny Novgorod, 603950, Russia }
\address {$^{73}$Pusan National University, Busan 609-735, Korea }
\address {$^{74}$Hanyang University, Seoul 133-791, Korea }
\address {$^{75}$University of Adelaide, Adelaide, South Australia 5005, Australia }
\address {$^{76}$Monash University, Victoria 3800, Australia }
\address {$^{77}$Seoul National University, Seoul 151-742, Korea }
\address {$^{78}$The Chinese University of Hong Kong, Shatin, NT, Hong Kong SAR, China }
\address {$^{79}$University of Alabama in Huntsville, Huntsville, AL 35899, USA }
\address {$^{80}$Southern University and A\&M College, Baton Rouge, LA 70813, USA }
\address {$^{81}$College of William and Mary, Williamsburg, VA 23187, USA }
\address {$^{82}$Instituto de F\'\i sica Te\'orica, University Estadual Paulista/ICTP South American Institute for Fundamental Research, S\~ao Paulo SP 01140-070, Brazil }
\address {$^{83}$University of Cambridge, Cambridge CB2 1TN, United Kingdom }
\address {$^{84}$IISER-Kolkata, Mohanpur, West Bengal 741252, India }
\address {$^{85}$Rutherford Appleton Laboratory, HSIC, Chilton, Didcot, Oxon OX11 0QX, United Kingdom }
\address {$^{86}$Whitman College, 345 Boyer Avenue, Walla Walla, WA 99362 USA }
\address {$^{87}$National Institute for Mathematical Sciences, Daejeon 305-390, Korea }
\address {$^{88}$Hobart and William Smith Colleges, Geneva, NY 14456, USA }
\address {$^{89}$King's College London, University of London, London WC2R 2LS, United Kingdom }
\address {$^{90}$Andrews University, Berrien Springs, MI 49104, USA }
\address {$^{91}$Trinity University, San Antonio, TX 78212, USA }
\address {$^{92}$University of Washington, Seattle, WA 98195, USA }
\address {$^{93}$Kenyon College, Gambier, OH 43022, USA }
\address {$^{94}$Abilene Christian University, Abilene, TX 79699, USA }
\address {$^{95}$Universit\`a degli Studi di Urbino ``Carlo Bo'', I-61029 Urbino, Italy and INFN, Sezione di Firenze, I-50019 Sesto Fiorentino, Italy}

%\author{M Evans$^{8}$} \email{mevans@ligo.mit.edu}
%\affiliation{Massachusetts Institute of Technology, Cambridge, MA 02139, USA \vup}

%\affiliation{INFN, Sezione di Firenze, I-50019 Sesto Fiorentino, Italy\\
%            Universit\`a degli Studi di Urbino ``Carlo Bo'', I-61029 Urbino, Italy \vup}

\date{\today}

\begin{abstract}
The second-generation of gravitational-wave detectors are just starting operation, and have already yielding their first detections.
Research is now concentrated on how to maximize the scientific potential of gravitational-wave astronomy.
To support this effort, we present 
here design targets for a new generation of detectors, which will be capable of observing 
compact binary sources with high signal-to-noise ratio throughout the Universe.
\end{abstract}

%\keywords{}

\maketitle


%%%%%%%%%%%%%%%%%%%%%%%%%%%%%%%%%%%%
% some local definitions
\def\Parm{P_{\rm arm}}
\def\Larm{L_{\rm arm}}
\def\rb{r_{\rm beam}}
\def\pavg{\phi_{\rm eff}}

\def\Msolar{M_{\odot}}

\def\letter{letter}

%%%%%%%%%%%%%%%%%%%%%%%%%%%%%%%%%%%%
\section{Introduction}
\sslabel{intro}

With the development of extremely sensitive ground-based \gw\ detectors
 \cite{aLIGO2015,Virgo2014,Kagra2013} and
 the recent detection of \gws\ by LIGO \cite{detection2016,PhysRevLett.116.241103},
 extensive theoretical work is going into understanding potential gravitational-wave (GW) sources
  \cite{lrr-2009-2,Sathya2012,Broeck2014,Kinugawa2015,Read2009,Messenger2012,DelPozzo2012,
  TheLIGOScientific:2016htt,Baiotti:2016qnr,Meacher:2015iua}.
In order to guide this investigation, and to help direct instrument research and development,
 in this \letter\ we present design targets for a new generation of detectors.

The work presented here builds on a previous study of how the fundamental noise sources
 in ground-based GW detectors scale with detector length \cite{Miller2015,Dwyer2015},
 and is complementary to the detailed sensitivity analysis of the Einstein Telescope
  (ET, a proposed next generation European detector) presented in \cite{Hild2011, ET-0106C-10}.
The ET analysis will not be reproduced in this work, but the ET-D sensitivity curve
 from \cite{Hild2011} is used for comparison.
It represents one 10\,km long detector consisting of two interferometers \cite{Hild10},
 the detector arms forming a right angle.
The ET design consists
 of three co-located detectors in a triangular geometry~\cite{Freise09},
 but for the purpose of this \letter\ we compare the sensitivity of single detectors,
 all with arms at right angles.
(A comparison of triangular
 and right angled detector sensitivities can be found in \cite{Freise2008}.)
 
%%%%%%%%
\MEfig{h!}{0.5}{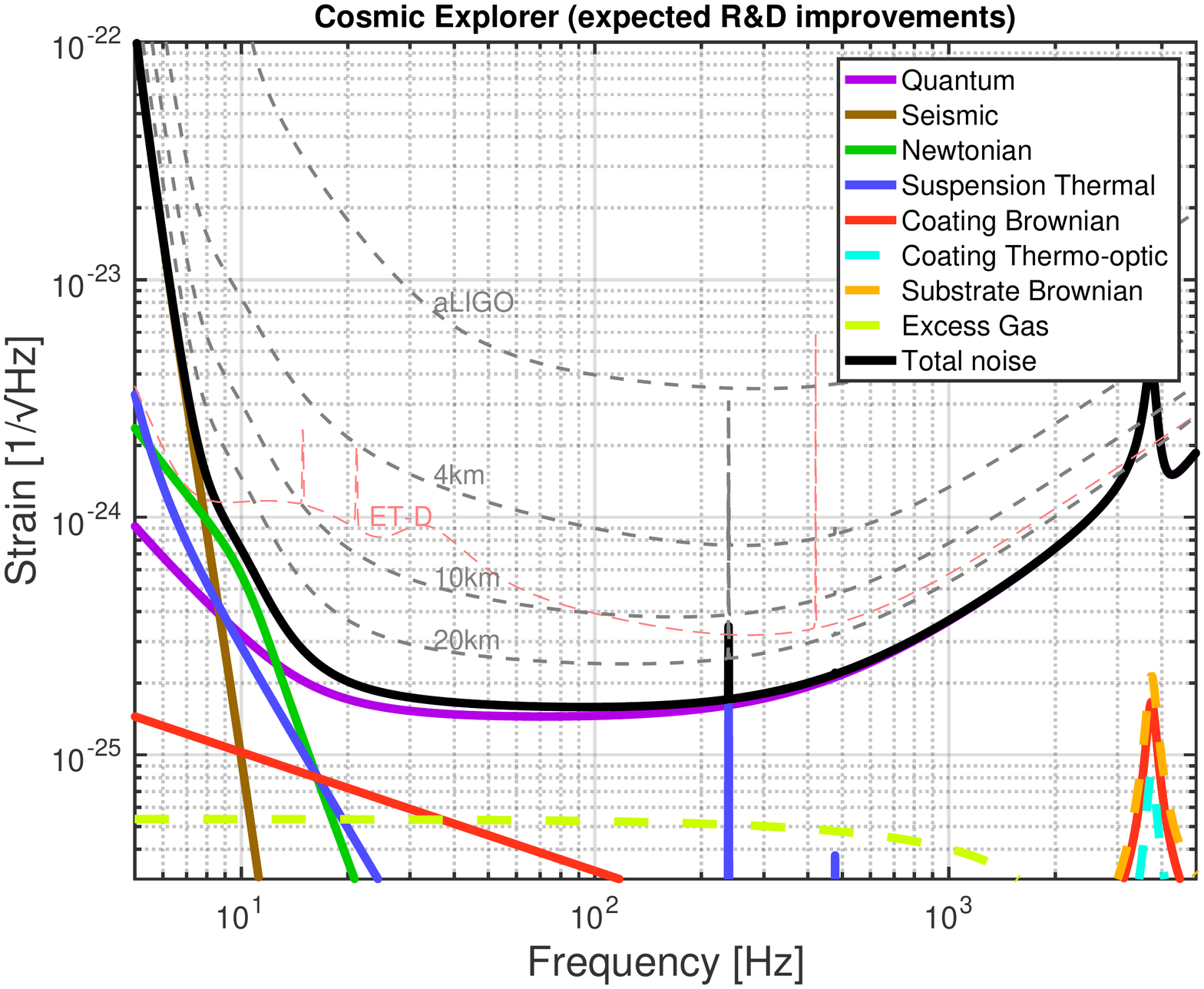}{target}{Target sensitivity for a next generation gravitational-wave detector,
 known as ``Cosmic Explorer'' for its ability to receive signals from cosmological distances.
The solid curves are for a \SI{40}{km} long detector,
 while the dashed grey curves show the sensitivity of shorter, but technologically similar detectors;
 lengths are 4, 10 and \SI{20}{km}.
 The Advanced LIGO and Einstein Telescope design sensitivities are also shown for reference.}
%%%%%%%%

From this work two important conclusions emerge.
The first of these is that the next generation of GW detectors will
 be capable of detecting compact binary sources with high signal to noise ratio (${\rm SNR} > 20$)
 even at high redshift ($z > 10$).
The second is that there are multiple distinct areas of on-going research and development (R\&D)
 which will play important roles in determining the scientific output of future detectors.

In what follows, we start by expressing the sensitivity of a next-generation GW detector
 as a collection of target values for each of the fundamental noise sources.
This is followed by discussions of the R\&D efforts that could plausibly attain
 these goals in the course of the next 10 years.
We conclude with a brief discussion of science targets, which will be
 accessible to a world-wide network of next-generation detectors.

%%%%%%%%%%%%%%%%%%%%%%%%%%%%%%%%%%%%%%%%%%%%%%%%%%%%%
\section{Next Generation Sensitivity}

The target sensitivity of a \SI{40}{km} long next generation GW detector,
 known as ``Cosmic Explorer'', is shown in figure \fref{target}.
The in-band sensitivity and upper end of the band,
 from \SI{10}{Hz} to a few kilohertz, is determined by quantum noise,
 while the lower limit to the sensitive band is determined by local gravitational disturbances
 (known as ``Newtonian noise'' or NN \cite{Harms2015}).
Other significant in-band noise sources are coating thermal noise and residual gas noise.
Seismic noise and suspension thermal noise, though sub-dominant, also serve to define
 a lower bound to the detector's sensitive band.
Each of these noise sources will be discussed in detail in the following sections.

The estimated sensitivities presented here are computed from analytical models of 
 dominant noises and interferometer response in the sensitive frequency band
 of the detector.
All of the contributing noise sources shown in figure \fref{target}
 are intended as targets that could plausibly be attained by 
 a number of on-going research programs,
 rather than curves linked to a particular technology.
As such, in each of the following sections we give simple scaling relationships,
 which show how these noises scale relative to the relevant parameters,
 along with the values used to produce the target curves.

%%%%%%%%%%%%%%%%%%
\subsection{Quantum Noise}

Laser interferometer based GW detectors are almost inevitably limited in their sensitivity by
 the quantum nature of light.
In most of the sensitive band, this limit comes in the form of counting statistics or
 ``shot noise'' in the photo-detection process.
Typically near the low-frequency end of the band a similar limit appears in the form of
  quantum radiation pressure noise (RPN), which can be thought of as the sum of impulsive
  forces applied to the interferometer mirrors as they reflect the photons incident upon them.
 A unified picture of quantum noise is, however,
  necessary to understand correlations between shot noise and radiation pressure noise
  and to appreciate the possibility of reducing quantum noise through the
  use of squeezed vacuum states of light
  \cite{Caves1985, Buonanno2001, McClelland2011,Schnabel2010}.

In this \letter\!\,, we use the now standard ``dual recycled Fabry-Perot Michelson'' interferometer (DRFPMI)
 configuration, which is common to all kilometer-scale second generation detectors
  \cite{Adhikari2014, Aso2013, aLIGO2015}.
While this choice is considered likely for the next generation of detectors,
 a number of plausible alternative designs are being actively investigated \cite{Miao2015, Voronchev2014,Somiya2016521,PhysRevD.87.096008,PhysRevD.86.062001,Liv.Rev.Rel.15.2012}.
 
For a DRFPMI, the optical response to GW strain is essentially determined by the choice of
 signal extraction cavity configuration
 \footnote{The term ``signal recycling'' is often used to refer to any interferometer
  configuration that uses a mirror at the output port of the interferometer to change
  the interferometer response.  However, more careful language distinguishes between 
  cases where this mirror \emph{decreases} the signal
  storage time in the interferometer, known as ``signal extraction'',
  and cases where it \emph{increases} the signal
  storage time in the interferometer, known as ``signal recycling''.}.
We will assume for simplicity a ``broadband signal extraction'' configuration, in which the
 signal extraction cavity is operated on resonance, and the detector bandwidth is set by the
 choice of signal extraction mirror reflectivity.
Figure \fref{wideband} shows the effect of increased signal extraction mirror reflectivity relative to
 that shown in figure \fref{target}; the detector bandwidth is somewhat wider, but the in-band sensitivity
 is reduced \cite{Buonanno2001,Harms2003,Miao2014}.

An important technology which will determine the quantum limited sensitivity of future
 GW detectors is squeezed light \cite{McClelland2011}.
Squeezed states of light have been demonstrated to be effective in reducing quantum noise
 in GW interferometers \cite{GEO_SQZ_2011,Barsotti2013}, and have been incorporated into the plans
 for all future detectors \cite{Hild2011, Miller2015}.
The impact of squeezing on the scientific output of GW detectors has been studied in
 detail in \cite{Lynch2015}.
In this analysis, we assume frequency dependent squeezing,
 as described in \cite{Evans2013, Kwee2014, Oelker2016a}.

For any given DRFPMI configuration choice, the quantum noise is determined by the power
 in the interferometer, the laser wavelength, the level of squeezing at the readout,
 and at low-frequencies (where radiation pressure noise is dominant)
 by the mass of the interferometer mirrors.
For any \emph{fixed detector bandwidth}, the in-band sensitivity scales as
\beq{qshot}
\frac{h_{\rm shot} }{h_{0\, \rm shot}}  = \sqrt{ \frac{\SI{2}{MW}}{\Parm}}
  \sqrt{  \frac{\lambda}{\SI{1.5}{\mu m}} } \( \frac{3}{r_{\rm sqz}} \)   \sqrt{ \frac{\SI{40}{km}}{\Larm} }
\eeq
\beq{qrad}
\frac{h_{\rm RPN} }{h_{0\, \rm RPN}} = \sqrt{ \frac{\Parm}{\SI{2}{MW}} }
 \sqrt{ \frac{\SI{1.5}{\mu m}}{\lambda} } \( \frac{3}{r_{\rm sqz}} \) \( \frac{\SI{320}{kg}}{m_{\rm TM}}\)
  \( \frac{\SI{40}{km}}{\Larm} \)^{3/2} \!\!\!\!, \nonumber
\eeq
where $\Parm$ is the circulating power in the arm cavities of length $\Larm$
 bounded by mirrors of mass $m_{\rm TM}$,
$\lambda$ is the laser wavelength and $r_{\rm sqz}$ is observed squeezing level
(e.g., $r_{\rm sqz} = 3$ corresponds to approximately a \SI{10}{dB} noise reduction).
The values normalizing each parameter in the above scaling relations
 are the ones used to produce the curves shown in figure \fref{target},
 such that the resulting ratio ($h_X / h_{0X}$) is relative to the target noise amplitude spectral density.
All of the values used to produce the target sensitivity curves are presented in table \ref{params}.

The exact choice of laser wavelength, for instance, is not important as long as
 longer wavelengths are accompanied by higher power.
%Similarly, changing the power or wavelength will not change the shape of the sensitivity curve
% if the mass of the optics are changed proportionately.
As an important example of this, consider two future interferometers;
 one uses fused silica optics and operates with \SI{1.4}{MW} of \SI{1064}{nm} light in the arms,
 while the other uses silicon optics and operates with \SI{2.8}{MW} of \SI{2}{\mu m} light in the arms.
Both interferometers will have essentially the same quantum noise.
%Similarly, doubling the power and the mirror mass will result in a $\sqrt{2}$ lower quantum noise at all frequencies.

Interestingly, quantum noise does not scale inversely with length.
This is due to the \emph{fixed detector bandwidth} constraint,
 which requires increased signal extraction with greater length to maintain a constant integration time.
While the shot noise appears to increase due to reduced signal gain in the interferometer,
 the radiation pressure noise is reduced (both relative to $1/L$).
A hidden dependence which is not included in equation \eref{qrad} is the dependence of
 the mirror mass $m_{\rm TM}$ on length; longer interferometers generally have larger beams
 and thus require larger and more massive mirrors.
 
There are several areas of R\&D which will determine the quantum noise in future detectors.
First among these is work into increasing the measured squeezing levels
 \cite{Isogai2013, Vahlbruch2008,Dwyer2013,Grote2013,Chua2014,Dooley2015,
 Vahlbruch2016,Oelker2016b,Wade2016,Schreiber2016}.
Second is prototyping of the alternative configurations to demonstrate suppression
 of quantum radiation-pressure noise at low frequencies \cite{2014_CQG.31.215009_Graef},
 and to investigate the influence of imperfections on this ability \cite{2015_NJP17.043031_asymSag}.
Less easily explored in tabletop experiments, but equally relevant,
 are thermal compensation, alignment control and parametric instabilities,
 which determine the maximum power level that can be used
 in an interferometer
 \cite{Waldman2006, Sidles2006, Evans:2015cu}.
Finally, the ability to produce and suspend large mirrors will be necessary for any next generation
 GW detector \cite{Hild2011,Cumming2012}, and will have a beneficial impact on low-frequency quantum noise.

%%%%%%%%
\MEfig{t!}{0.5}{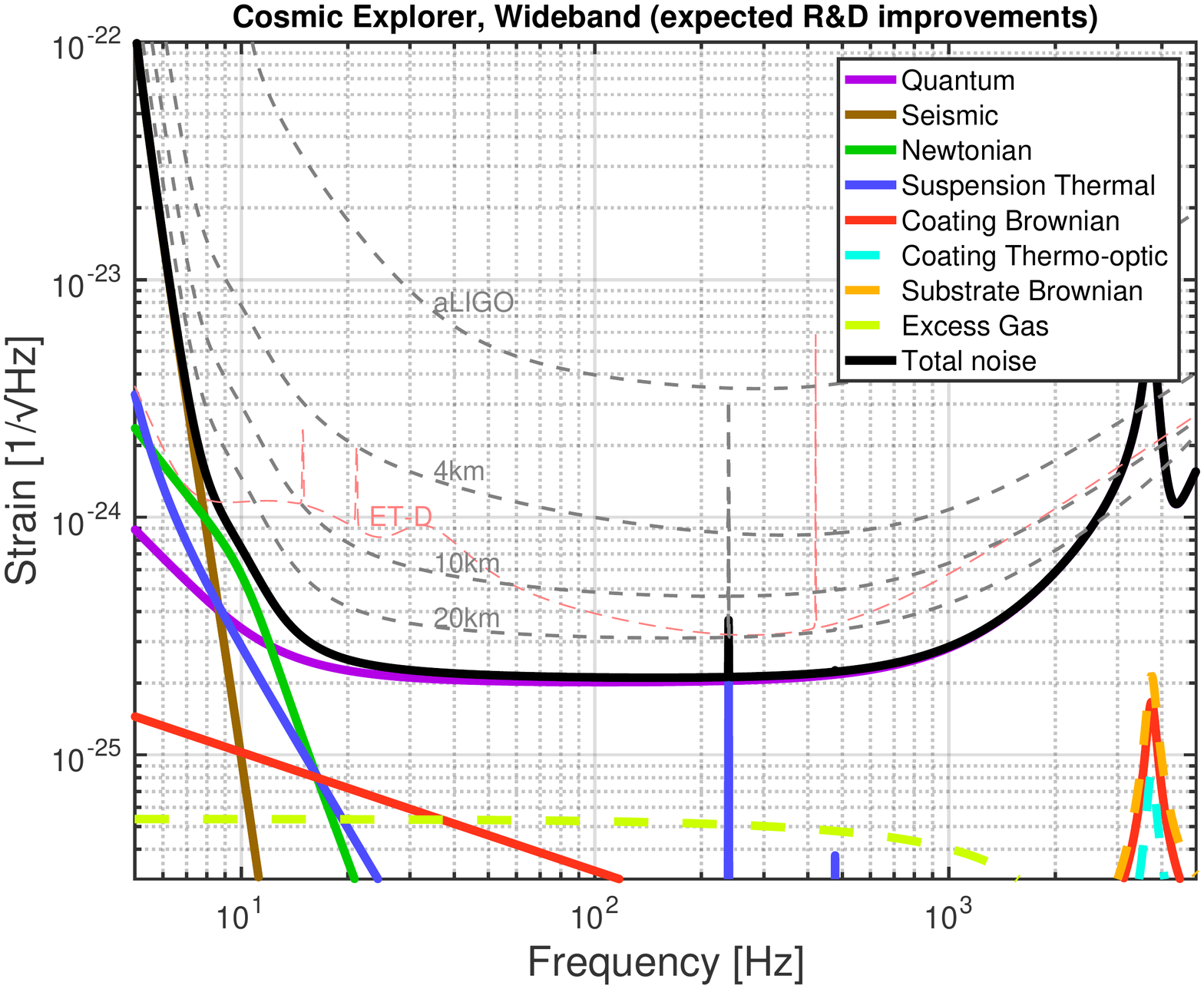}{wideband}{Similar to figure \fref{target} but with a more reflective
 signal extraction mirror which gives a wider sensitive band, but is less sensitive in-band.
The tradeoff between in-band sensitivity and bandwidth will need to be optimized
 to maximize specific science objectives
 (e.g., testing general relativity with black hole binaries,
  measuring neutron star equation of state, detection of GW from supernovae, etc.).
The dashed grey curves show the sensitivity of shorter, but technologically similar detectors;
 lengths are 4, 10 and \SI{20}{km}.}
%%%%%%%%

%%%%%%%%
\MEfig{t!}{0.5}{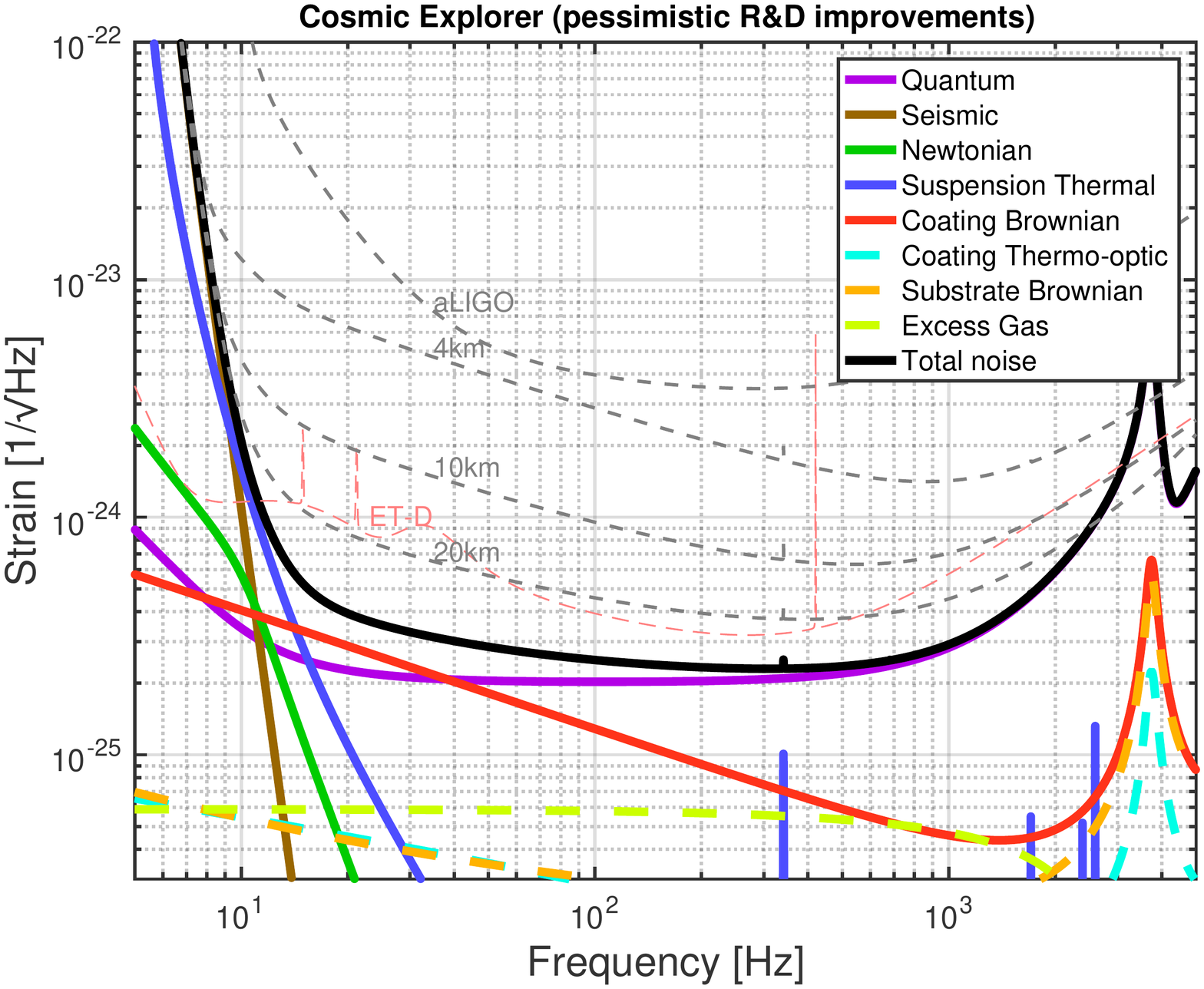}{pessimistic}{Similar to figure \fref{wideband}
 but with coating and suspension thermal noise models which assume minimal progress.
The wide-band signal extraction choice is made to minimize the impact of CTN.
The proximity of the dashed grey \SI{4}{km} curve to the Advanced LIGO reference curve
 reflects the fact that coating technology, which is nearly limiting in Advanced LIGO,
 becomes dominant over a range of frequencies given the reduction of quantum noise
 assumed for the future.}
%%%%%%%%

%%%%%%%%%%%%%%%%%%
\subsection{Coating Thermal Noise}

Coating thermal noise (CTN) is a determining factor in GW interferometer designs;
 in current (second generation) GW detectors, CTN equals quantum noise in the most
 sensitive and most astrophysically interesting part of the detection band around \SI{100}{Hz}
  \cite{Adhikari2014, Hong2013, Yam2015}.
For instance, the Advanced LIGO  detectors were designed to minimize
 the impact of CTN by maximizing the laser spot sizes on the mirrors
 (at the expense of alignment stability in the interferometer),
 and the Kagra detector design is dominated by the incorporation
 of cryogenics to combat thermal noise \cite{Somiya2012,Aso2013}.
Similarly, current R\&D into cryogenic technologies for future detectors is largely driven by the
 need to reduce CTN, either directly through low-temperature operation,
 or indirectly through changes in material properties as a function of temperature.

Holding all else constant, CTN scales as
 \beq{ctn}
\frac{h_{\rm CTN} }{h_{0\, \rm CTN}}  = \sqrt{ \frac{T}{\SI{123}{K}}}
  \sqrt{  \frac{\pavg}{5 \times 10^{-5}}}  \( \frac{\SI{14}{cm}}{\rb} \)
  \( \frac{\SI{40}{km}}{\Larm} \),
\eeq
 where $T$ is the temperature, $\pavg$ is volume- and direction-averaged
 mechanical loss angle of the coating (defined below in equation \eref{phiavg}),
 and $\rb$ the beam size on the interferometer mirrors
 ($1/e^2$ intensity).
Thus, the brute-force techniques to reducing CTN are lowering the temperature and
 increasing the beam radius, while finding low-loss materials
 is an active and demanding area of research.
  
 To be precise, $\pavg$ is the effective mechanical loss angle of the coating,
 \beq{phiavg}
% \pavg = S_{\rm CTN} \frac{(2 \pi \times \SI{30}{Hz})(\pi \rb^2)}{4 k_B T ~ \sum_j d_j}
% \frac{Y_s}{1 - \sigma_s - 2 \sigma_s^2}
 \pavg = \frac{\sum_j b_j d_j \phi_{Mj}}{2 \sum_j d_j}
 \eeq
   in the notation of equation 1 in \cite{Yam2015},
   where the summations run over all coating layers,
   $d_j$ is the layer thickness, $\phi_{Mj}$ is the mechanical loss angle,
   and $b_j$ is a factor of order unity which depends on the mechanical properties of
   the substrate and coating (numerically, $b_j \!\sim\! 2$ for most coatings).
 This is related to $h_{0\, \rm CTN}$ by (again in the notation of \cite{Yam2015})
 \beq{phiavg}
h_{0\, \rm CTN}^2 = \frac{8 k_B T (1 - \sigma_s - 2 \sigma_s^2)}{ \pi \rb^2 \Larm^2 \omega Y_s}
 \pavg \sum_j d_j,
 \eeq
 where the summation gives the total coating thickness summed over all four test-mass mirrors
  (for the target design this is $16.6 \lambda$).

 It should be noted that a number of important dependencies are hidden in equation \eref{ctn}.
 In particular, $\pavg$ may have a strong dependence on $T$,
  and for a fixed cavity geometry $\rb$ grows with $\Larm$ such that
 \beq{ctn2}
\frac{h_{\rm CTN} }{h_{0\, \rm CTN}}  = \sqrt{ \frac{T}{\SI{123}{K}}}
  \sqrt{  \frac{\pavg(T)}{5 \times 10^{-5}}}  \( \frac{\SI{40}{km}}{\Larm} \)^{3/2}
\eeq
 is an equally valid scaling relation.
Along the same lines, both $\rb$ and the coating thickness grow with $\lambda$,
 but they do so such that the effects cancel for fixed cavity geometry and finesse.
 
While the CTN curves in figures \fref{target} and \fref{wideband} are based on plausible
 extrapolations from current lab-scale results \cite{Cole2016, Steinlechner2016},
 figure \fref{pessimistic} shows a family of sensitivity curves which
 assume little or no progress is made in reducing CTN.

%%%%%%%%%%%%%%%%%%
\subsection{Newtonian Noise}

The motion of mass from seismic waves or atmospheric pressure and temperature changes produce local gravitational disturbances, which couple directly to the detector and cannot be distinguished from gravitation waves \cite{Sau1984,Cre2008,Harms2015}. The power spectrum of such disturbances, known as ``Newtonian noise'' (NN), is calculated to fall quickly with increasing frequency, such that while it presents a significant challenge below \SI{10}{Hz}, it is negligible above \SI{30}{Hz}. The level of NN present in a given detector is determined by the facility location (e.g., local geology, seismicity and weather) and construction (e.g., on the surface or underground), and defines the low-frequency end of the sensitive band for that facility. 

Active research in the area of NN will determine important aspects of the design of future GW detector facilities. Feed-forward cancellation of ground motion NN using a seismometer array has shown the potential to provide some immunity \cite{Cel2000,DHA2012,Harms2015}, whereas concepts for feed-forward cancellation of atmospheric perturbations still need to be developed. It is also the case that the spectrum of atmospheric infra-sound and wind driven NN is, as yet, poorly understood and cancellation appears more challenging than for seismic NN \cite{Harms2015}. Ongoing characterization of underground sites will also determine the gain for GW detectors with respect to NN reduction \cite{HaEA2010,BBR2015}, as future GW detectors may need to be constructed a few hundred meters underground if the sensitive band is to be extended below \SI{10}{Hz}.

An important aspect of site characterization is to estimate the effectiveness of a NN cancellation system, which above all depends on the distribution of local sources, and for sub-\SI{10}{Hz} detectors also on the complexity of local topography \cite{CoHa2012}.

Research in this area is developing quickly, and the NN estimates presented in this \letter\ assume a factor of 10 cancellation of seismic NN

\begin{table}
\begin{tabular}{c|c|c|c|c}
   & ~~~~CE~~~~ & ~~CE pess~~  & ~ET-D (HF)~  & ~ET-D (LF) \\
   \hline
 $\Larm$        & \SI{40}{km}       &  \SI{40}{km}  & \SI{10}{km}       & \SI{10}{km} \\
 $\Parm$        & \SI{2}{MW}        &  \SI{1.4}{MW}& \SI{3}{MW}         &  \SI{18}{kW} \\
 $\lambda$  & \SI{1550}{nm}     & \SI{1064}{nm}& \SI{1064}{nm}  & \SI{1550}{nm} \\
 $r_{\rm sqz}$  & 3                 & 3             & 3                 & 3 \\
 $m_{\rm TM}$ & \SI{320}{kg}        & \SI{320}{kg}  & \SI{200}{kg}      & \SI{200}{kg} \\
 $\rb$      & \SI{14}{cm}       & \SI{12}{cm}   &  \SI{9}{cm}       &  \SI{7}{cm} ($\rm LG_{33}$)\\
 $T$            & \SI{123}{K}       & \SI{290}{K}   & \SI{290}{K}       & \SI{10}{K} \\
 $\pavg$        & $5 \times 10^{-5}$ & $ 1.2 \times 10^{-4}$ & $ 1.2 \times 10^{-4}$ & $1.3 \times 10^{-4}$  \\
 \hline
\end{tabular}
\caption{Parameters used to produce the Cosmic Explorer (CE) target curve.
 The CE pessimistic and Einstein Telescope, high- and low-frequency (HF and LF)
  parameters are included for comparison.\label{params}}
\end{table}

%%%%%%%%%%%%%%%%%%
\subsection{Suspension Thermal Noise and Seismic Noise}

Suspension thermal noise and seismic noise,
 particularly in the direction parallel to local gravity (``vertical''),
 can place an important limit on the low-frequency sensitivity of future GW detectors \cite{Hammond2012}.
This is true both because, like NN, this noise source falls quickly with increasing frequency,
 but also because the coupling of vertical motion to the sensitive direction of the GW detector
 increases linearly with detector length (due to the curvature of the Earth),
 making the GW strain resulting from a fixed vertical displacement
 noise level insensitive to detector length \cite{Dwyer2015}.

Current research into test-mass suspensions is focused on supporting larger masses
 (required by detectors with $\Larm > \SI{10}{km}$), and longer suspensions for reduced thermal
 and seismic noise both in the horizontal and vertical directions \cite{Hammond2012}.
Vertical thermal noise can be further reduced by lowering the vertical resonance frequency
 of the last stage of the suspension,
 possibly by introducing monolithic blade springs into the suspension designs \cite{Cumming2012}.

%%%%%%%%%%%%%%%%%%
\subsection{Residual Gas Noise}

Gravitational wave detectors operate in ultra-high vacuum to avoid phase noise due to
 acoustic and thermal noise that would make in-air operation impossible.
The best vacuum levels in the long-baseline arms of current detectors are near
$\SI{4 \times 10^{-7}}{Pa} \simeq \SI{3 \times 10^{-9}}{torr}$
 and are dominated by out-gassing of $\rm H_2$ from the beam-tube steel.
This noise scales with average laser-beam cross-section and arm length as \cite{ZuWh1996}

 \beq{res}
\frac{h_{\rm gas} }{h_{0\, \rm gas}}  =  \sqrt{ \frac{p_{\rm gas}}{\SI{4 \times\! 10^{-7}}{Pa}} } ~
  \sqrt{ \frac{\SI{14}{cm}}{r_{\rm beam}} } ~ \sqrt{ \frac{\SI{40}{km}}{\Larm} }.
\eeq

%%%%%%%%%%%%%%%%%%%%%%%%%%%%%%%%%%%%
\section{Compact Binaries at high red-shift\\ and Extragalactic Supernovae}
\sslabel{science}
% !TEX root = FutureGWD.tex

The high sensitivity of future ground-based gravitational wave detectors will considerably expand their
 scientific output relative to existing facilities.
Clearly, sources routinely detected already by current instruments in the local universe will be
 detected frequently with high SNR, and at cosmological distances. 
Straightforward examples are binary systems involving black holes and neutron stars.
These systems, referred to collectively as ``compact binaries'' (CBCs), are ideal GW emitters and a rich
 source of information about extreme physics and astrophysics, which is inaccessible by other means
 \cite{lrr-2009-2,Sathya2012,Broeck2014,Kinugawa2015,Read2009,Gair:2010dx,Baiotti:2016qnr}. 

Binary neutron stars (BNS) could yield precious information about the equation of state (EOS) of neutron stars, which can complement or improve what can be obtained with electromagnetic radiation~\cite{2014ApJ...787..136P,2012ARNPS..62..485L}.
However, second-generation detectors would need hundreds of BNS detections to distinguish between competing EOS~\cite{2013PhRvL.111g1101D,2015PhRvD..92b3012A,2015PhRvD..91d3002L}. New detectors would help both by providing high SNR events, and increasing the numbers of threshold events~\cite{2010PhRvD..81l3016H}.

In general, all studies that rely on detecting a large numbers of events will benefit from future detectors.
Examples include estimating the mass and spin distribution of neutron stars and black holes in binaries, as well as their formation channels~\cite{2015ApJ...807L..24L,2015ApJ...810...58S,2015arXiv150304307V}.

Furthermore, a GW detector with the sensitivity shown in figure \fref{target} could detect a significant fraction of binary neutron star systems even at $z = 6$, during the epoch of reionization, beyond which few such systems are expected to exist \cite{1538-3881-122-6-2850}.
Those high-redshift systems could be used to verify if BNS are the main producer of metals in the Universe~\cite{2016Natur.531..610J}, and as standard candles for cosmography \cite{Messenger2012}.

%%%%%%%%
\MEfig{t!}{0.5}{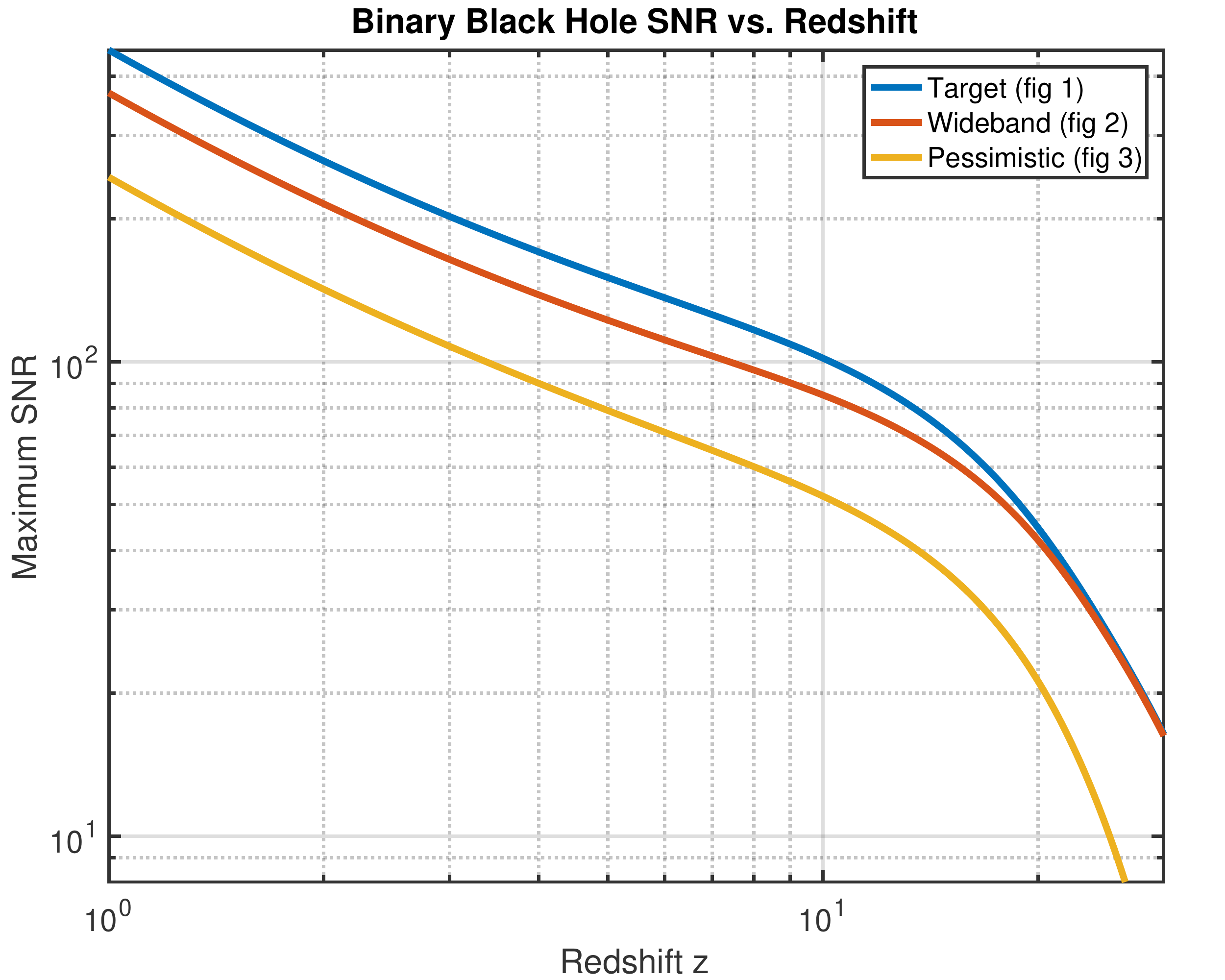}{snr}{The maximum signal-to-noise ratio (SNR) for which GW detectors
 with the sensitivities shown in figures \fref{target}, \fref{wideband} and \fref{pessimistic} 
 would detect a system made of two black holes (each with an intrinsic mass \SI{30}{\Msolar}),
  as a function of redshift.
Many systems of this sort will be detected at $z < 2$ with an ${\rm SNR} > 100$,
 enabling precision tests of gravity under the most extreme conditions.}
%%%%%%%%

Future instruments could detect a system made of two \SI{30}{\Msolar} black holes,
 similar to the first system detected by LIGO \cite{detection2016},
 with a signal-to-noise ratio of 100 at $z = 10$,
 thus capturing essentially all such mergers in the observable universe
  (see figure \fref{snr}).
  
Nearby events would have even higher SNRs, allowing for exquisite tests of general relativity~\cite{GW150914-TESTOFGR}, and measurements of black-hole mass and spins with unprecedented precision.
The possibility of observing black holes as far as they exist could give us a chance to observe the remnants of the first stars, and to explore dark ages of the Universe, from which galaxies and large-scale structure emerged.
 
Furthermore, future detectors may be able to observe GW from core-collapse supernovae, whose gravitational-wave signature is still uncertain~\cite{,Fryer:2011zz,2016arXiv160501785A}. GWs provide the only way to probe the interior of supernovae, and could yield precious information on the explosion mechanism. Significant uncertainty exists on the efficiency of conversion of mass in gravitational-wave energy, but even in the most optimistic scenario the sensitivity of existing GW detectors to core-collapse supernovae is of a few megaparsec~\cite{2016PhRvD..93d2002G}. 
A factor of ten more sensitive instruments could dramatically change the chance of positive detections. In fact, while the rate of core-collapse supernovae is expected to be of the order of one per century in the Milky Way and the Magellanic clouds, it increases to $\sim 2$ per year within 20~Mpc~\cite{2012ApJ...756..111M,2012A&A...537A.132B}.

%%%%%%%%%%%%%%%%%%%%%%%%%%%%%%%%%%%%
\section{Conclusions}
\sslabel{Conclusions}

We present an outlook for future gravitational wave detectors and how their
 sensitivity depends on the success of current research and development efforts.
While the sensitivity curves and contributing noise levels
 presented here are somewhat speculative,
 in that they are based on technology which is expected to be operational
 10 to 15 years from now, they represent plausible targets for the next
 generation of ground-based gravitational wave detectors.
By giving us a window into some of the most extreme events in the Universe,
 these detectors will continue to revolutionize our understanding of both fundamental physics
 and astrophysics.

%%%%%%%%%%%%%%%%%%
\begin{acknowledgments}
The authors would like to acknowledge the invaluable wisdom derived from
 interactions with members of the Virgo and Kagra collaborations
 without which this work would not have been possible.
 
LIGO was constructed by the California Institute of Technology and
 Massachusetts Institute of Technology with funding from the National Science Foundation,
 and operates under cooperative agreement PHY-0757058.
Advanced LIGO was built under award PHY-0823459.
This paper carries LIGO Document Number LIGO-P1600143.
\end{acknowledgments}

%%%%%%%%%%%%%%%%%%%%%%%%%%%%%%%%%%%%%%%%%%%%%%%%%%%%%
%%%%%%%%%%%%%%%%%%%%%%%%%%%%%%%%%%%%%%%%%%%%%%%%%%%%%


\bibliographystyle{apsrev}
\bibliography{papers}

\end{document}